# Dynamical noisy canalization in morphogenesis: lessons from *Hydra* regeneration

Oded Agam[1]* and Erez Braun[2]*

Developmental robustness is framed as progress through a fixed Waddington-type landscape. We argue that in morphogenesis this landscape evolves through coupled bio-signaling, mechanical, and physiological processes, while fluctuations aid exploration. In *Hydra* regeneration, stochastic $Ca^{2+}$ activity plays a major role in reshaping the landscape of accessible morphologies as regeneration unfolds, including the early progressive confinement of tissue fluctuations. We propose testing this framework of dynamical noisy canalization in developmental systems.

## Introduction

How do reproducible body forms emerge from tissues whose molecular activity, physiological state, and mechanical properties fluctuate over time? Developmental robustness is commonly described via Waddington's canalization, in which trajectories proceed through an epigenetic landscape toward stable outcomes [1-3]. In many applications, this landscape is treated as fixed: development unfolds through pre-existing terrain[4-7]. Morphogenesis, however, typically involves a separation of timescales. Fast shape changes occur under constraints set by slower processes, including signaling, mechanics, electrical activity, tissue infrastructure, and other physiological dynamics that co-evolve with the tissue[8-16].

This observation calls for a dynamical canalization framework in morphogenesis. Its central object is a landscape of accessible morphologies: the tissue configurations that are dynamically available, likely to be explored, or stabilized at a given stage. As development or regeneration proceeds, this evolved set of accessible shapes reflects coupled physiological, mechanical, and biochemical processes that restrict and bias the forms a tissue can explore. Dynamic landscapes are well established for cell-fate stabilization and reprogramming, where valleys are created, reshaped, or eliminated as parameters change[17-20]. Here we extend this logic to morphogenesis, where the relevant coordinates are defined by the tissue morphological configurations.

*Hydra* regeneration provides a tractable realization of this process. Small tissue fragments fold into nearly spherical shells and regenerate into viable animals whose head-foot polarity is biased by positional information inherited from the parent[10, 21-23]. Yet, the process remains sensitive to physiological and electrical perturbations[13, 24, 25]. Modulating the tissue $Ca^{2+}$ activity can halt regeneration, reverse a fully formed morphology, and under combined perturbations, restore it[13, 26-28]. This degree of control over morphological dynamics points to an underlying landscape that is actively reshaped as regeneration proceeds[29].

We coin this process *dynamical noisy canalization* (Fig. 1): canalization because the spectrum of shapes available to the tissue progressively narrows; dynamical because that landscape changes over time; and noisy because fluctuations aid exploration and transitions. *Hydra* thus provides an experimentally accessible case for a broader open question: how self-organizing developmental tissues access and stabilize a reproducible form, while reshaping the very landscape that guides them.

## Halting, reversing, and restoring *Hydra* regeneration

Heptanol is a gap-junction blocker that suppresses $Ca^{2+}$ activity in the epithelial bilayer and arrests *Hydra* regeneration above a threshold concentration[26, 30-33]. Below threshold, regeneration proceeds normally: the initial sphere-like fragment elongates, forms a head, tentacles and a foot, and matures. This occurs mainly by reorganizing cells already present at the onset, with effectively no cell division[25, 34, 35]. At higher Heptanol concentrations, the tissue remains nearly spherical for days, without apparent damage. The arrest is reversible upon washout.

[1] The Racah Institute of Physics, The Hebrew University of Jerusalem, Jerusalem 9190401, Israel.
[2] Department of Physics and Network Biology Research Laboratories, Technion-Israel Institute of Technology, Haifa 32000, Israel.
* Correspondence should be addressed to O.A (agam.oded@gmail.com ) or E.B. (erezph1@gmail.com)

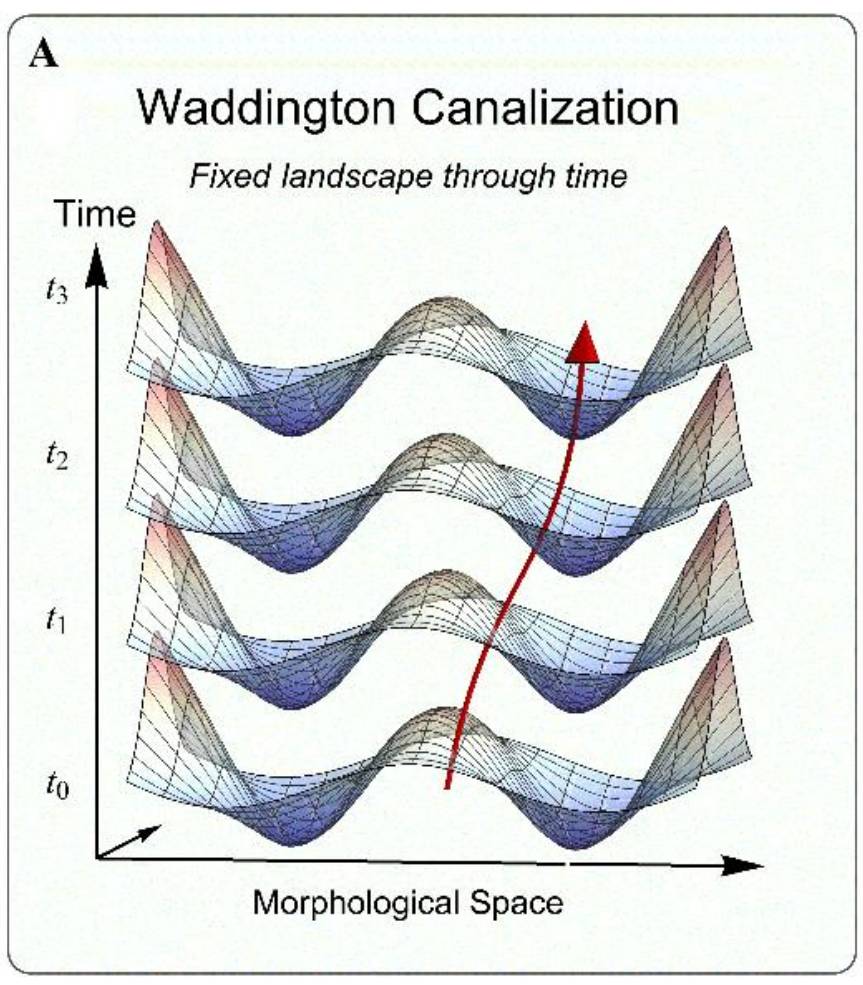

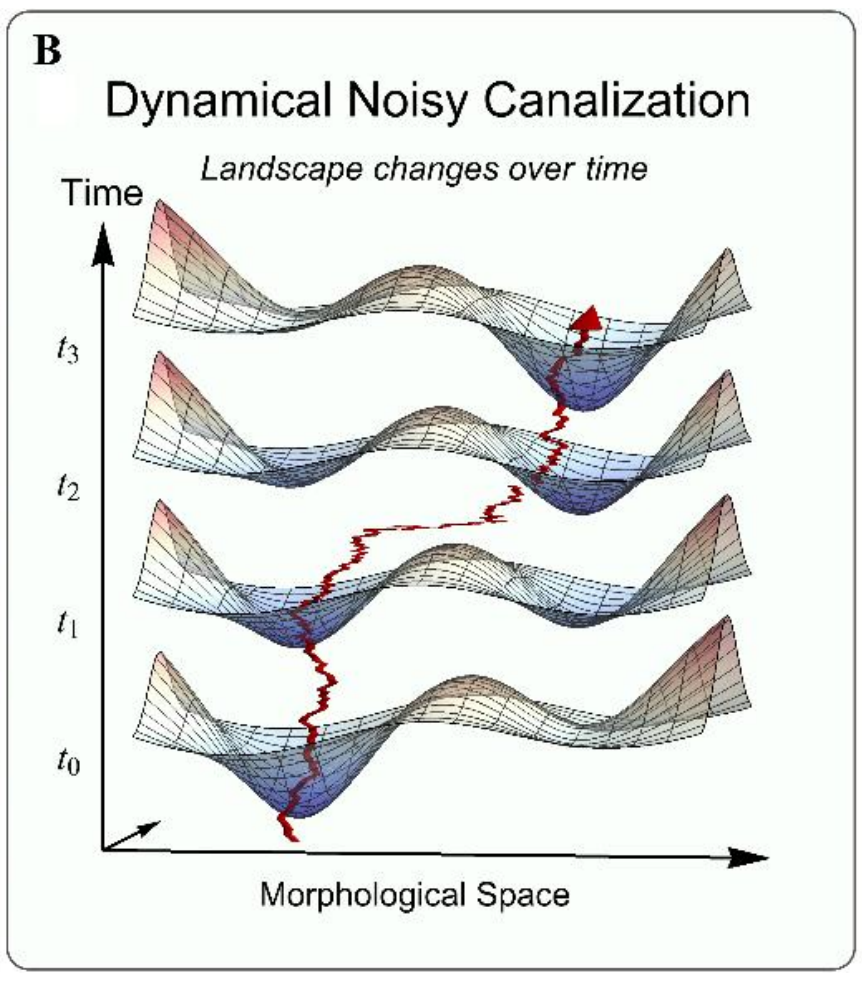

**Fig. 1. Fixed versus evolving developmental landscapes.** Two views of canalization in morphogenesis, shown as landscapes over a schematic morphological space at successive times $t_0$ to $t_3$. **A**, Waddington's classical picture: the landscape is fixed, and development proceeds as a trajectory through pre-existing valleys. Robustness arises from the geometry of the landscape itself. **B**, Dynamical noisy canalization: the landscape evolves over time. The tissue trajectory (red) is stochastic, and fluctuations contribute both to shaping the landscape and to driving transitions across barriers. In *Hydra* regeneration, the early landscape favors a spherical configuration (deep well on the left at $t_0$); over time, the landscape tilts and reshapes until the elongated configuration (right well) becomes the stable outcome. Robustness here is a property of the coupled dynamics between tissue and landscape, not of a pre-existing terrain.

A constant electric field produces another form of arrest. Above a threshold field strength, regeneration stops and the tissue remains nearly spherical[13], even though $Ca^{2+}$ activity is enhanced rather than suppressed[36].

Strikingly, either perturbation can also reverse morphology after complete regeneration: tentacles retract and the body returns back to a nearly spherical state[13]. This dramatic back-and-forth modulation can be repeated several times[13, 27].

The most revealing experiment combines the two perturbations. A fragment held under Heptanol above threshold concentration, can still regenerate after more than 60 hours of arrest if an electric field is applied simultaneously with the drug. Under this treatment, $Ca^{2+}$ activity recovers, and regeneration resumes[26]. Each perturbation alone halts regeneration, yet together they restore it. Both Heptanol and the electric field likely affect the tissue in multiple ways, but the rescue of morphogenesis by opposing effects on $Ca^{2+}$ activity strongly points to a patterned $Ca^{2+}$ field as a relevant morphogenetic variable.

The arrest, reversal, and rescue sequence is difficult to accommodate within any standard framework on its own. Canonical bio-signaling, including Wnt/β-catenin[37-43], reaction-diffusion models[44], tissue stretching[42, 43, 45], and actomyosin nematic order[10, 22], each capture essential aspects of *Hydra* morphogenesis. Yet none directly explains why suppressing $Ca^{2+}$ activity, enhancing it in a way that disrupts its spatiotemporal patterning, and then combining the two perturbations should produce arrest, reversal, and rescue. The experiment points to a level of description in which structured $Ca^{2+}$ activity is not merely a downstream marker, but an observable component of the dynamics that determines which shapes are available to the tissue.

The framework developed below rests on two interlinked pillars: the $Ca^{2+}$ field within the tissue and the evolving landscape of accessible shapes. The experiments track tissue shapes and the $Ca^{2+}$ activity in fragments excised from defined positions along the parent axis, using simultaneous bright-field and $Ca^{2+}$ fluorescence imaging.

## The calcium field

The $Ca^{2+}$ in the *Hydra* is a spatiotemporal active field, shaped by electrical excitability and coupled to the physiological state of the regenerating tissue[13, 33, 36, 46, 47]. Three features make it central to our framework.

First, the $Ca^{2+}$ field carries inherited polarity information. In tissue fragments excised from a mature *Hydra*, a persistent spatial $Ca^{2+}$ gradient appears early in regeneration, well before visible morphogenesis begins[28]. This bias reflects polarity inherited from the parent rather than a pattern generated entirely de novo. The gradient correlates with the future body plan: regions of relatively high and low average $Ca^{2+}$ activity correspond to the future head and foot, respectively[28].

Second, local $Ca^{2+}$ activity is stochastic[36]. The persistent gradient is a slow average over excitable events: localized domains, much smaller than the tissue, rapidly switch between low-activity and excited states. This behavior appears during normal regeneration, under electric field, under Heptanol, and under mechanical confinement[26, 36].

Third, the $Ca^{2+}$ field helps maintain tissue integrity. During regeneration, an osmotic pressure gradient inflates the epithelial shell until local ruptures release fluid; the tissue then repairs the rupture, and the cycle resumes[42, 45, 48, 49]. The rupture size, reflected in the tissue's volume modulation, depends on $Ca^{2+}$-regulated actomyosin contractility around the rupture site[46]. In normally regenerating tissue, large ruptures are strongly suppressed. When $Ca^{2+}$ ion entry into the tissue is disrupted, this control is compromised and large ruptures become more likely[46]. The bounded rupture-size distribution is therefore an actively regulated state involving $Ca^{2+}$-dependent contractility.

Taken together, these observations point to a central role for the $Ca^{2+}$ field in *Hydra* morphogenesis. Yet this field does not by itself specify global shape. The next question is how it constrains the configurations available to the regenerating tissue fragment.

**The morphological potential**

The main morphological transition in *Hydra*, from a sphere-like form to a cylindrical body, can unfold over minutes[26, 27]. This is too rapid to be explained primarily by cell division, gene expression, or large-scale tissue rearrangements. On this timescale, shape is governed mainly by tissue mechanics. $Ca^{2+}$ activity affects these mechanics by triggering actomyosin contractions, which are balanced by tissue elasticity and hydrostatic pressure in the enclosed cavity.

The relation between $Ca^{2+}$ activity and shape is bidirectional. $Ca^{2+}$ activity drives actomyosin contractions that reshape the tissue, while local tissue deformations bias the location and strength of the $Ca^{2+}$ activity[26, 28]. Thus, different $Ca^{2+}$ activity patterns can generate different contraction patterns and favor different tissue configurations. Experimental observations show that $Ca^{2+}$ excitations occur preferentially in flatter regions rather than sharply curved ones, a relationship captured in our model by coupling $Ca^{2+}$ activity to local curvature[26].

We call the resulting coarse-grained landscape underlying tissue shapes the *morphological potential* defined over morphological space[26]. Because the $Ca^{2+}$ field is stochastic, shapes are not sampled with equal probabilities: frequently visited configurations lie low in this potential, rarely visited ones lie high, and transitions depend on the barriers between them[27]. The morphological potential summarizes how $Ca^{2+}$ activity, tissue mechanics, hydrostatic pressure, and other underlying variables restrict the morphologies dynamically available to the regenerating fragment[50].

This view allows reinterpretation of the arrest-and-rescue experiments. Under Heptanol, $Ca^{2+}$ activity is strongly suppressed, active contractile forces are weakened relative to the elastic and pressure-driven contributions[36], and the cylindrical configuration is no longer dynamically accessible on the observed timescale (Fig. 2B). Under a strong electric field, $Ca^{2+}$ activity is enhanced, but its persistent spatial organization is disrupted[13, 26, 27, 36]. Because local excitation domains enter refractory periods, increased firing rate excites many different locations over time. On the timescale of the tissue mechanical processes, $Ca^{2+}$ activity becomes effectively uniform, so contractile forces fail to favor elongation. The tissue again remains near the spherical configuration, but for a different reason (Fig. 2C).

The combined perturbation then becomes intelligible. Applying the electric field in the presence of Heptanol restores sufficient $Ca^{2+}$ activity and enough spatial structure to break symmetry. The morphological potential again supports an elongated configuration, allowing transition toward a mature body form (Fig. 2D). Thus, Heptanol, strong electric field, and their combination change which morphologies are dynamically available.

**Dynamical noisy canalization**

The morphological potential is not a fixed background through which the tissue proceeds. It is shaped along the regeneration process (Figs. 1B & 2A). In small fragments, excised from the central region of the parent Hydra, the tissue remains broadly quasi-spherical for tens of hours, with little change in its mean spherical-like shape, before transitioning to an elongated cylindrical form in only a few minutes[26]. These early dynamics are described as motion between the two configurations along a single reduced coordinate on which the potential behaves as a tilted double well[26, 27]. At first, the spherical well is favored and the tissue remains confined near this state. Over time, the elongated minimum deepens relative to the spherical one, until a fluctuation carries the tissue across the barrier between them (Fig. 2A).

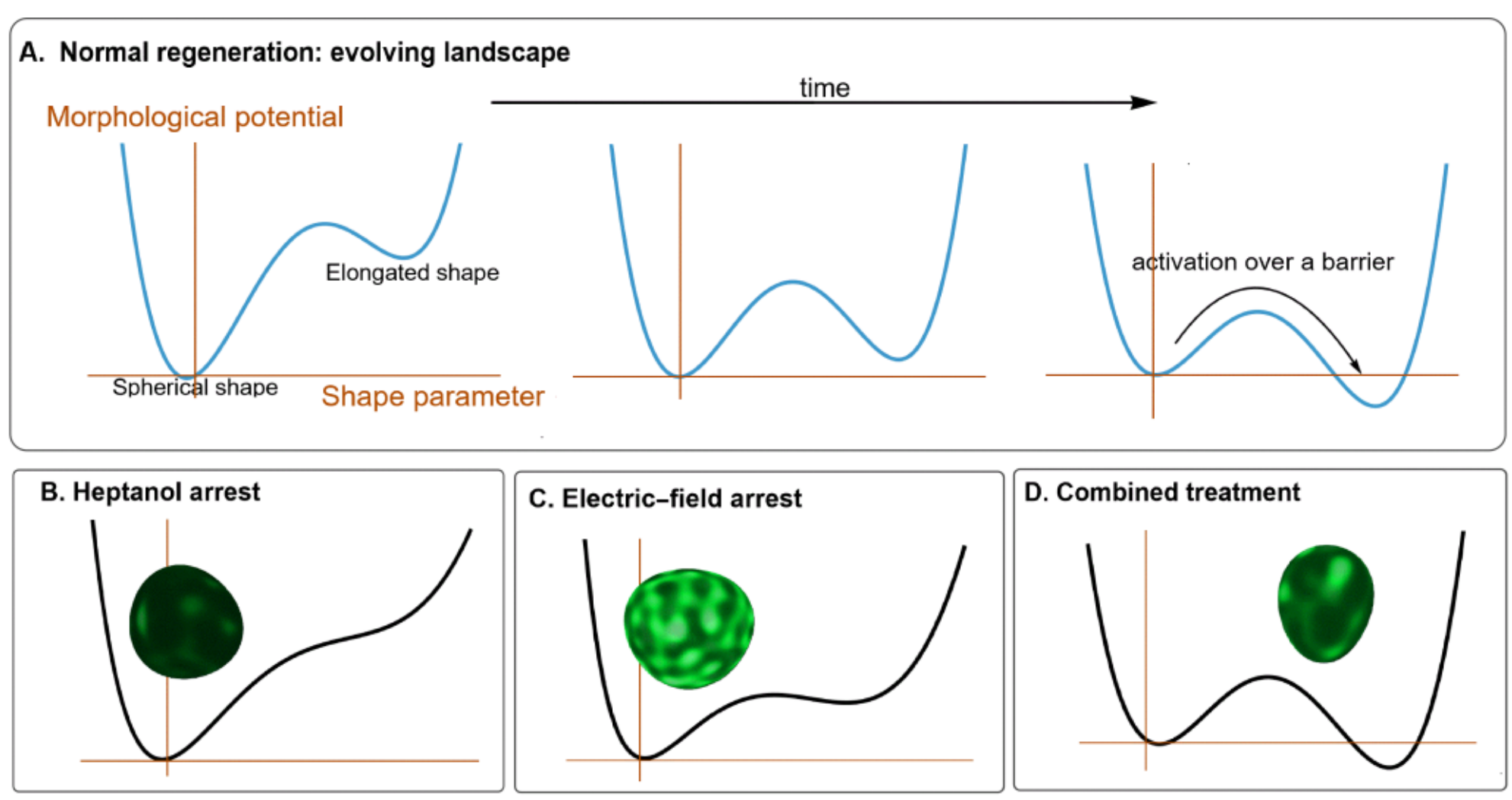

**Fig. 2. The morphological potential and its perturbations.** Schematic representation of the morphological potential along a shape coordinate ranging from a quasi-spherical configuration (left) to an elongated, cylindrical-like configuration (right). **A**, Evolution of the potential during normal regeneration. Early in regeneration (left), the potential has a single dominant well at the spherical configuration. As regeneration proceeds (center), a second well appears at the elongated configuration and the two wells become comparable in depth, separated by a barrier. Eventually (right), the elongated well becomes the deeper minimum, and the tissue crosses the barrier through fluctuation-driven activation. The visible sphere-to-cylinder transition occurs at this late stage but reflects landscape reshaping that has been ongoing throughout. **B-D**, Schematic morphological potential under three experimental conditions, together with illustrations of the characteristic $Ca^{2+}$ activity patterns in each case (insets). **B**, Heptanol arrest: $Ca^{2+}$ activity is suppressed, contractile forces are weak, and the cylindrical configuration is not dynamically accessible. **C**, Electric-field arrest: $Ca^{2+}$ activity is enhanced but spatially disorganized, and the resulting contractile forces are effectively uniform; the cylindrical configuration is again unfavored, but for a different reason than in B. **D**, Combined treatment: $Ca^{2+}$ activity is restored and acquires a spatial organization that allows contractile forces to break symmetry; the double-well structure is recovered, and the elongated configuration becomes the stable outcome.

This two-state structure is exposed when the dynamics are slowed by an appropriately chosen electric field[27]. Instead of making a single transition, the tissue undergoes repeated stochastic swings between near-spherical and elongated configurations before finally committing to the cylindrical form. These swings reveal the two principal minima of the potential. The gradual increase in residence time in the elongated state further indicates that the corresponding minimum is being lowered relative to the spherical one.

A further signature comes from stochastic resonance[51]. In a double-well potential, weak fluctuations rarely carry the system over a barrier, whereas very strong fluctuations make the transitions almost random[52]. At an intermediate noise level, a weak periodic drive can most efficiently bias barrier crossing[51]. This behavior is observed in regenerating *Hydra* tissue under weak periodic electric-field modulation[27]. Samples with intermediate fluctuation levels show the strongest response, and the inferred barrier is comparable to the effective noise strength estimated from spontaneous shape fluctuations. Thus, natural fluctuations are large enough to participate in transitions between the two morphological states.

Together, the rapid sphere-to-cylinder transition after a long waiting period, the stochastic morphological swings, and stochastic resonance support a coarse double-well structure of the morphological potential[26, 27]. A fourth observation reveals what this coarse picture does not capture: even near one state, the local structure of the potential can change. During the long interval before visible elongation, the tissue fragment remains broadly quasi-spherical, yet its shape fluctuation modes evolves[29]. Early on, the tissue explores many deformation modes; later, fluctuations are confined to a smaller set of collective modes. This restriction is captured by the effective dimension of fluctuation space, which estimates how many deformation modes contribute substantially to the dynamics[29]. The effective dimension decreases before the main morphological transition. Blocking $Ca^{2+}$ channels with $GdCl_3$ [53, 54] broadens the spectrum of accessible fluctuations, indicating that $Ca^{2+}$ activity helps constraining the deformation modes explored during this apparently quiescent period[29].

In morphological-potential terms, the minimum supporting the quasi-spherical configuration may remain nearly fixed in position while its local structure changes. The potential becomes more restrictive: some deformation directions are suppressed, whereas others remain available. While its mean morphology changes little, the fluctuation space around it is reorganized. The visible sphere-to-cylinder transition is not the onset of morphogenesis, but the late outcome of a potential that has been reshaped over time, both locally by constraining fluctuations as well as globally by tilting dynamics. This is the sense in which *Hydra* regeneration exhibits dynamical noisy canalization[55-58]: the tissue does not move through a fixed landscape[3, 4]; it progressively reorganizes its accessible fluctuation space until the elongated state becomes available and stable.

## Three conceptual shifts

Dynamical noisy canalization suggests three shifts in the description of morphogenesis. It does not replace molecular signaling, inherited polarity, or tissue mechanics, but treats them as coupled ingredients leading to an evolving landscape through which the tissue progressively restricts the morphological configurations it can explore.

The first shift concerns biochemical activity. In the classical morphogen picture, signal concentrations originating from a source (e.g., the head organizer) specify positional information that affects the local tissue or cell fates[59, 60]. In *Hydra* regeneration, an early $Ca^{2+}$ gradient does carry polarity information correlated with the future head-foot axis[28], yet $Ca^{2+}$ acts beyond positional encoding. It is a spatiotemporal excitable field that controls contractility[26, 27, 33], shapes the mechanical forces that deform the tissue fragment[26, 29, 50], and participates in tissue repair[46]. The shift is in the nature of the signal: from a positional map to an active field that participates in determining which morphogenetic trajectories are available to the tissue.

The second shift concerns the developmental landscape. Waddington's metaphor captures robustness since different trajectories converge to the same outcome[26, 29, 50]. *Hydra* regeneration adds that the landscape is also constructed by the dynamics it guides; it requires closed-loop dynamics[8] between the evolving tissue and the constraints acting on it. Canalization becomes the progressive reshaping of the landscape as the accessible fluctuation space narrows from the earliest stages.

The third shift concerns noise. Rather than treating noise as something robustness must tolerate, dynamical noisy canalization emphasizes the constructive role of noise generated within the tissue itself. In *Hydra*, $Ca^{2+}$ and shape fluctuations help carry the tissue across barriers between morphological states, as seen in stochastic swings[27], and participate in the restriction of accessible deformation modes[29]. Protein folding offers a conceptual analogy. Thermal fluctuations drive local exploration within a structured energy landscape whose funnels and barriers guide the molecule to a reproducible folded state[61]. *Hydra* morphogenesis may work similarly, with fluctuations channeled toward a reproducible body form. It differs in one important respect: the landscape itself is reshaped during the process, so the analogy is closer to intrinsically disordered proteins, which shift among structures depending on their environment and binding partners[62]. Comparable constructive roles for fluctuations appear across biological levels, from gene-regulatory circuits [63-65] to bistable cell-fate determination[66].

Together, these shifts recast our central question. Rather than asking how a molecular pattern specifies a body plan, we ask how an excitable field participates in structuring fluctuations and reshaping the space of accessible morphologies, so that many microscopic histories converge on a reproducible form.

## Relation to existing frameworks

The morphological potential offers a meeting point for established common frameworks of *Hydra* regeneration, describing how signaling, mechanics, physiological activity, and inherited positional information are integrated, reshaping the configurations available to the tissue as regeneration unfolds.

Mechanics enters most directly. Supracellular actomyosin fibers provide the substrate through which $Ca^{2+}$-regulated contractility acts on shape, while their organization carries morphological information of its own. Their nematic pattern is partly inherited and reorganizes during regeneration[10, 22]. Defects in this pattern act as organization centers, with local strain focused at defect sites and morphological features developing there[22, 23, 45]. In the morphological-potential picture, this organization helps determine which $Ca^{2+}$-dependent contractions remain local and which are converted into tissue-scale deformation modes, including axial elongation[26, 67].

Biochemical signaling enters as a set of controls on the same landscape. Organizer and reaction-diffusion frameworks identify molecular cues of axial identity, chiefly Wnt-dependent head-organizer activity and its inhibition[37-40, 42]. In the present view, such cues may bias the $Ca^{2+}$ field, tune excitability, or modify local mechanics, changing the forms available to the tissue[21, 26, 28].

Our experiments on positional memory illustrate this integration. A fragment's morphological trajectory depends on its origin along the parent's axis[28]. Central fragments switch most sharply between the quasi-spherical and elongated configurations, while fragments excised from regions nearer the head or foot, follow slower paths with longer-lived intermediates. Second-cut twin fragments derived from the same large, closed tissue shell follow more similar trajectories than unrelated fragments, indicating that part of the inherited bias persists after recutting and renewed morphogenesis[28].

The early $Ca^{2+}$ gradient may be one link between inherited polarity and later shape changes. With the early emergence of a foot precursor in our experiments, it points to head-foot control distributed across the tissue, rather than set by a single organizer source[28]. The morphological potential thus adds a temporal synthesis: it describes how these ingredients reshape the accessible morphological space during regeneration.

## Experimental expectations

Dynamical noisy canalization makes a central prediction: before the tissue commits to a visible morphological change, its accessible fluctuations should narrow, and its physiological activity should resolve toward the pattern that later drives the change. In *Hydra*, this appears as a decrease in the effective dimension of shape-fluctuation space prior to the sphere-to-cylinder transition[29], with $Ca^{2+}$ activity increasingly coupled to the most significant deformation modes, especially those associated with the future body axis[26]. If this principle is general, the same pre-morphological canalization should appear in other tissues that self-organize from weakly patterned initial states, e.g., in embryogenesis, in organ formation and in other regeneration systems like Planaria.

Synthetic systems such as mammalian gastruloids and stem-cell aggregates offer another natural test. They break symmetry and elongate from near-isotropic starting conditions, with outcomes shaped by size, signaling, and timing[68, 69]. Successful single-axis formation should be preceded by the relevant activity fields resolving toward one dominant axis, whereas multi-axis outcomes should show earlier signatures of competing activity foci or deformation modes.

A sharper test in *Hydra* itself is regeneration from dissociated cell aggregates, where tissue-level polarity is initially absent[25]. Successful regeneration should follow an organized $Ca^{2+}$ polarity field prior to a significant morphological transition: a single axis should be preceded by one dominant gradient, while multiple heads should be preceded by several competing polarity domains. The spatial structure of this field should matter at least as much as its mean level, so perturbations that preserve average $Ca^{2+}$ activity while altering its pattern (e.g., by optogenetics) should redirect the morphological trajectory[28, 36].

## Outlook: a dynamical layer in *Hydra* morphogenesis

$Ca^{2+}$ activity, together with the morphological potential, offer a dynamical layer that links polarity, mechanics, and physiological activity in time, without displacing the molecular, mechanical, or Waddingtonian accounts it draws on.

The evidence is strongest during early regeneration, when $Ca^{2+}$ activity and tissue shape can be followed simultaneously. The sphere-to-cylinder transition is therefore the first experimentally accessible instance of a broader process: an evolving landscape in which some configurations become more accessible, likely, or stable as regeneration proceeds. Later events, including foot and head formation, tentacle emergence, and possible cell-state changes, may be experimentally harder to follow but expected to involve the same principle in a higher-dimensional morphological space. The open challenge here is to follow $Ca^{2+}$ organization, tissue deformation, mechanics, and cell-state dynamics into these later stages.

Another central challenge is to determine how far the observed morphogenetic reversibility under $Ca^{2+}$ modulation extends. If reversal dynamics reach beyond morphology into biochemical patterning or cell differentiation, this would show that the dynamical layer is coupled deeply to the regeneration process.

More broadly, morphogenesis can be viewed as the progressive organization of the space of possible forms, in which $Ca^{2+}$ activity, mechanics, polarity, and signaling act as coupled processes rather than a separate set of instructions. Dynamical noisy canalization provides a language for this view: robustness arising through fluctuations within an evolving landscape.

**Acknowledgment**

We thank Kinneret Keren, Shani Maoz and Shimon Marom for useful comments on the manuscript.

**Funding**

This work was supported by a joint grant (OA & EB) from the Israel Science Foundation (Grant No. 1586/25).

**Author contributions**

O. Agam and E. Braun wrote the main manuscript text. O. Agam prepared figures. Both authors reviewed the manuscript.

**Competing interests**

The authors declare no competing interests

**Ethics, Consent to Participate, and Consent to Publish declarations:** not applicable.